# Murine AI excels at cats and cheese: Structural differences between human and mouse neurons and their implementation in generative AIs

Rino Saiga[1], Kaede Shiga[1]*, Yo Maruta[1]*, Chie Inomoto[2], Hiroshi Kajiwara[2], Naoya Nakamura[2], Yu Kakimoto[3], Yoshiro Yamamoto[4], Masahiro Yasutake[5], Masayuki Uesugi[5], Akihisa Takeuchi[5], Kentaro Uesugi[5], Yasuko Terada[5], Yoshio Suzuki[6], Viktor Nikitin[7], Vincent De Andrade[7], Francesco De Carlo[7], Yuichi Yamashita[8], Masanari Itokawa[9,10], Soichiro Ide[9], Kazutaka Ikeda[9,11], and Ryuta Mizutani[1,12,+]

[1]Department of Bioengineering, Tokai University, Hiratsuka, Kanagawa 259-1292, Japan
[2]Department of Pathology, Tokai University School of Medicine, Isehara, Kanagawa 259-1193, Japan
[3]Department of Forensic Medicine, Tokai University School of Medicine, Isehara, Kanagawa 259-1193, Japan
[3]Department of Cell Biology, Tokai University School of Medicine, Isehara, Kanagawa 259-1193, Japan
[4]Department of Mathematics, Tokai University, Hiratsuka, Kanagawa 259-1292, Japan
[5]Japan Synchrotron Radiation Research Institute (JASRI/SPring-8), Sayo, Hyogo 679-5198, Japan
[6]Photon Factory, High Energy Accelerator Research Organization KEK, Tsukuba, Ibaraki 305-0801, Japan
[7]Advanced Photon Source, Argonne National Laboratory, Lemont, IL 60439, USA
[8]Department of Information Medicine, National Institute of Neuroscience, National Center of Neurology and Psychiatry, Kodaira, Tokyo 187-8502, Japan
[9]Tokyo Metropolitan Institute of Medical Science, Setagaya, Tokyo 156-8506, Japan
[10]Tokyo Metropolitan Matsuzawa Hospital, Setagaya, Tokyo 156-0057, Japan
[11]Department of Neuropsychopharmacology, National Institute of Mental Health, National Center of Neurology and Psychiatry, Kodaira, Tokyo 187-8553, Japan
[12]SPring-8 Center, RIKEN, Sayo, Hyogo 679-5148, Japan

*These authors equally contributed to this work.
+Correspondence: mizutanilaboratory@gmail.com

**Abstract**

Mouse and human brains have different functions that depend on their neuronal networks. In this study, we analyzed nanometer-scale three-dimensional structures of brain tissues of the mouse medial prefrontal cortex and compared them with structures of the human anterior cingulate cortex. The obtained results indicated that mouse neuronal somata are smaller and neurites are thinner than those of human neurons. These structural features allow mouse neurons to be integrated in the limited space of the brain, though thin neurites should suppress distal connections according to cable theory. We implemented this mouse-mimetic constraint in convolutional layers of a generative adversarial network (GAN) and a denoising diffusion implicit model (DDIM), which were then subjected to image generation tasks using photo datasets of cat faces, cheese, human faces, and birds. The mouse-mimetic GAN outperformed a standard GAN in the image generation task using the cat faces and cheese photo datasets, but underperformed for human faces and birds. The mouse-mimetic DDIM gave similar results, suggesting that the nature of the datasets affected the results. Analyses of the four datasets indicated differences in their image entropy, which should influence the number of parameters required for image generation. The preferences of the mouse-mimetic AIs coincided with the impressions commonly associated with mice. The relationship between the neuronal network and brain function should be investigated by implementing other biological findings in artificial neural networks.

**Introduction**

Human and mouse brains differ in a number of aspects. The most obvious one is their size. The human brain is characterized by its folded cerebral cortex, of which the area is estimated to be 1500–2000 cm$^2$ (Van Essen & Drury, 1997; Im et al., 2008; Kang et al., 2012; Schnack et al., 2015). Moreover, the human cerebral cortex is typically 1.5–3.5 mm thick (Fischl & Dale, 2000; Schnack et al., 2015; Wagstyl et al. 2020) and consists of 10–20 billion neurons (Larsen et al., 2006; Herculano-Houzel, 2009; von Bartheld et al., 2016). Hence, the mean areal density of neurons in the human cerebral cortex is estimated to be approximately $10^7$ neurons/cm$^2$. In contrast, mice have tiny brains with a smooth cerebral cortex having a thickness of typically 1 mm (Pagani et al., 2016; Hammelrath et al., 2016) and an area of approximately 2 cm$^2$. The mean areal density of neurons in the mouse cerebral cortex is reported to be $9.3 \times 10^6$ neurons/cm$^2$ (Keller et al., 2018). This value nearly coincides with that of the human brain, though the cortical thickness differs by a factor of 2 to 3 between human and mouse.

These anatomical differences between human and mouse brains should originate from their cellular makeup. Comparative studies on the microscopic or cellular constitution of human and mouse brains have been reported from a wide variety of perspectives and have revealed similarities and differences between them (Tecott, 2003; Yáñez et al., 2005; Strand et al., 2007; Katzner & Weigelt, 2013; Hodge et al., 2019; Benavides-Piccione et al., 2020; Szegedi et al., 2020; Bakken et al., 2021; Beauchamp et al., 2022; Loomba et al., 2022; Kim et al., 2023; Ragone et al., 2024). A study on brain-wide transcriptomic data on humans and mice indicated that sensorimotor brain areas exhibit a higher degree of similarity than supramodal areas (Beauchamp et al., 2022). Indeed, the visual cortexes of humans and mice have been reported to share structural and functional principles (Katzner & Weigelt, 2013). Single-nucleus RNA-sequencing analysis of the temporal cortex revealed a well-conserved cellular architecture between humans and mice (Hodge et al., 2019). Transcriptomic and epigenomic profiling of cells in the primary motor cortex also showed a broadly conserved cellular composition (Bakken et al., 2021). However, it is not fully understood how the cellular makeup is translated into functional differences of the brain between humans and mice.

Analyses of the relationships between the neuronal network and brain function can provide clues to understanding the functional basis of the brain. We have recently reported nanometer-scale three-dimensional studies of human cerebral tissues of the anterior cingulate cortex and the superior temporal gyrus of schizophrenia and control cases (Mizutani et al., 2019, 2021, 2023). The results indicated that the neurites of schizophrenia cases are significantly thinner and more tortuous than those of controls and that the neurite curvature in the anterior cingulate cortex showed a correlation with the auditory hallucination score (Mizutani et al., 2023). Since hallucinations have been reported to be associated with the structure and connectivity of the anterior cingulate cortex (Scarfo et al., 2024; Panikratova et al., 2023; Chen et al., 2024), micro-structural differences of neurons observed in this area should be relevant to mental function.

Findings from studies on real neuronal networks have been using in designing artificial neural networks (Hassabis et al., 2017; Macpherson et al., 2021). The most successful example is the convolutional layer (Fukushima, 1980; Homma et al., 1988) that was inspired from studies on the visual cortex of mammal brains (Hubel & Wiesel, 1962). Since sensorimotor areas including the visual cortex show similarities between humans and mice (Katzner & Weigelt, 2013), biological findings regarding those areas would be informative for artificially reconstructing brain functions common to most animals. However, functions specific to humans cannot be reproduced from analyses of commonalities between humans and other animals. Neuronal networks in brain areas other than the sensorimotor cortex should be investigated in order to find the network characteristics responsible for higher brain functions.

In this study, we analyzed the three-dimensional structures of brain tissues of the mouse medial prefrontal cortex by using the synchrotron radiation X-ray nanotomography (nano-CT). Three-dimensional images of brain tissues of nine mice were traced in order to build Cartesian-coordinate models of their neuronal networks. The obtained models were used to evaluate structural parameters including the shape of neuronal somata and the spatial trajectory of neurites, which were then compared with the results of the human anterior cingulate cortex (Mizutani et al., 2019, 2023). The observed structural characteristics of mouse neurons were then translated into connection constraints and implemented in generative AIs in order to investigate their computational consequences. We examined the results in light of the network characteristics as well as the statistical nature of datasets used for training the generative AIs.

**Results**

**Three-dimensional structures of mouse brain tissues**

Brain tissue structures of layer V of the medial prefrontal cortex of nine mice (M1–M9) were visualized with nano-CT (Supplementary Table S1), in the same manner as previously reported for the corresponding layer of the anterior cingulate cortex of eight human control cases (Mizutani et al., 2019, 2023). Tomographic slices were reconstructed from the obtained X-ray images and stacked to reproduce three-dimensional structures (Figure 1A, Supplementary Figures S1–S20). The tissue structures were traced in order to build Cartesian-coordinate models (Figure 1B), as was previously performed for human cases (Figure 1C, D). The traced structures of the nine mice consisted of 109 neurons, 5665 neuronal processes, and 65541 dendritic spines in total (Supplementary Table S2–S4). The tissue structures of the eight human control cases consisted of 150 neurons, 4597 neuronal processes and 18559 spines (Mizutani et al., 2019, 2023). Differences between the human and mouse structures are discernible in Figure 1, such as in the shape of neuronal somata and the spatial trajectory of neurites, which indicates that human and mouse cerebral tissues can be distinguished simply by visualizing their structures.

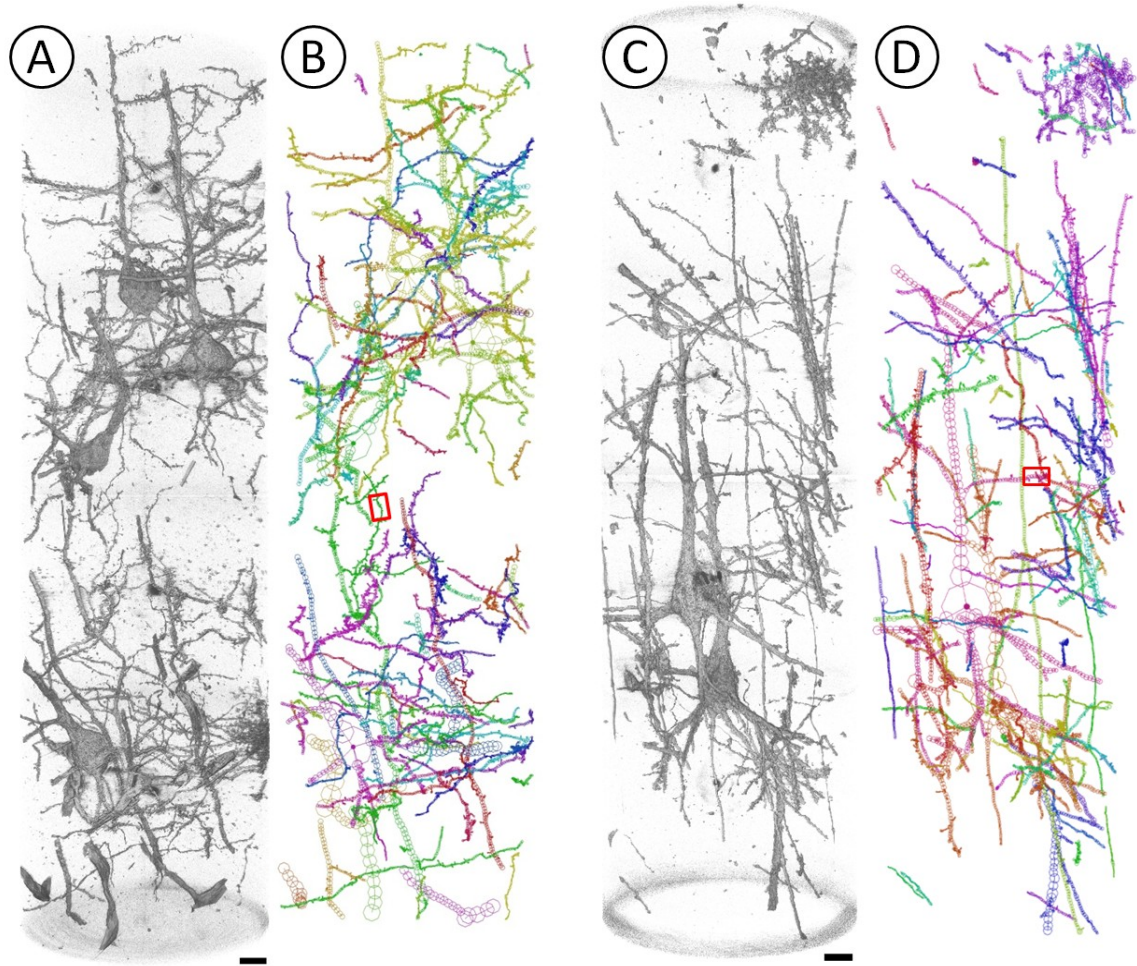

**Figure 1.** Three-dimensional structures of cerebral tissues of mouse and human. Brain tissues were stained with the Golgi method and visualized with nano-CT. Scale bars: 10 μm. (**A**) Rendering of mouse M8A structure (Supplementary Table S4) of the medial prefrontal cortex. Voxel values 40–800 were rendered in a perspective view by using the scatter HQ algorithm of the VG Studio Max software (Volume Graphics, Germany). Pial surface is toward the top. (**B**) Cartesian coordinate model of M8A drawn in a parallel view. Structures in the M8A image were traced in order to reconstruct them in Cartesian coordinate space by using the MCTrace software (Mizutani et al., 2013). Tissue constituents are color-coded. Nodes composing the model are drawn as octagons. The red box indicates the neurite shown in Figure 3A. (**C**) Rendering of human N8A structure of the anterior cingulate cortex. Reproduced from Mizutani et al., 2023. (**D**) Cartesian coordinate model of N8A. The red box indicates the neurite shown in Figure 3B.

**Structural differences between human and mouse brain tissues**

The human and mouse brain tissues (Figure 1) show differences in their neuronal soma shape. The somata of the mouse neurons are nearly spherical, while those of human are vertically triangular. Another difference is in their neurite structures. Mouse neurites are tortuous and thin, while human neurites are rather straight and thick (Figure 1).

Figure 2 summarizes differences in the soma shape. The most significant difference is in the neuronal soma length, which was defined as the length along spherical nodes having a diameter larger than half the diameter of cell soma (Figure 2A). Mean soma length for mouse is less than 60% that of human (Figure 2B; $p = 1.10 \times 10^{-6}$, two-sided Welch's *t*-test, Holm–Bonferroni corrected, same hereafter unless otherwise noted). The mouse soma width, which was defined with the diameter of the largest spherical node in the neuron, is approximately 85% of the human soma width (Figure 2B; $p = 0.025$). The shape difference is discernible from three-dimensional renderings of the neuronal somata (Figure 2E). The mouse soma is small and spherical, and hence compatible with thin and areally-confined cortex of the mouse brain. The human soma is long and wide, and hence conforms to the human cerebral cortex with sufficient depth and area.

The soma length difference was further examined by classifying neurons into pyramidal neurons and interneurons. Here, the difference was highly significant for pyramidal neurons (Figure 2C; $p = 6.3 \times 10^{-8}$). In contrast, the difference in interneuron length was insignificant (Figure 2D), though interneurons are less abundant and were identified only in five mouse and three human cases. These results indicated that the structural difference between human and mouse soma observed in this study is ascribable to pyramidal neurons.

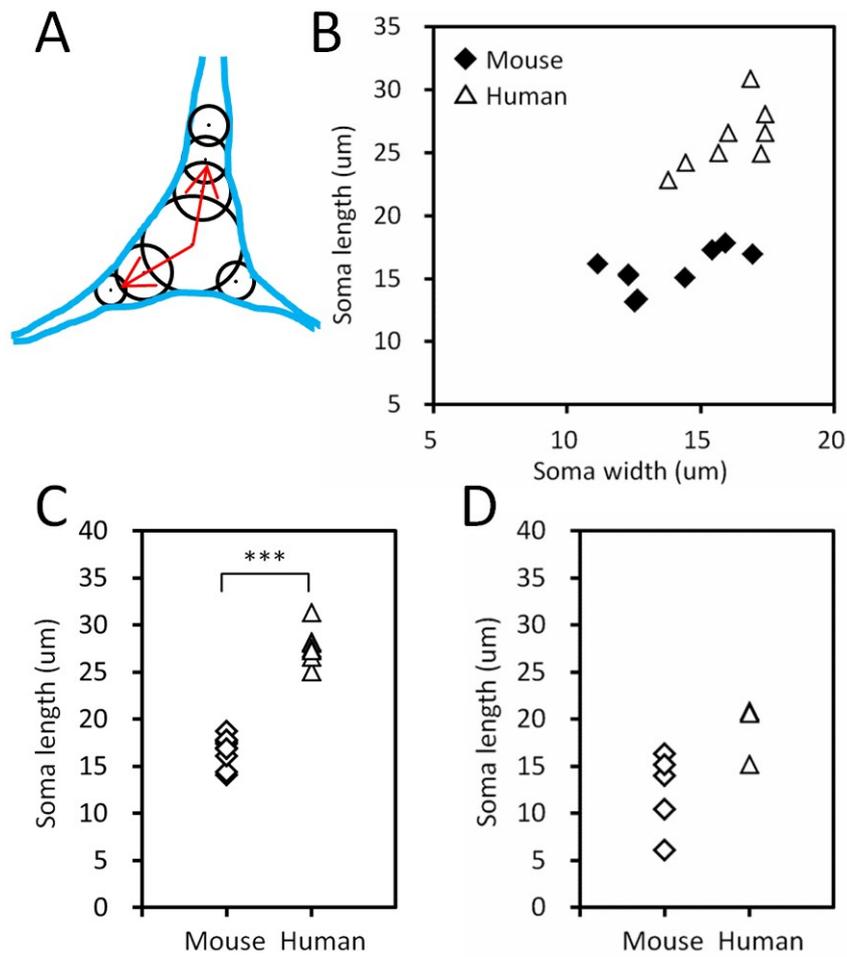

**Figure 2.** Differences in soma between human and mouse. (**A**) The soma width was defined with the diameter of the largest spherical node. The soma length was defined as the length along nodes having a diameter larger than half the soma width (red arrow). (**B**) Scatter plot of mean soma length and mean soma width. Mouse data are indicated with diamonds and human data with triangles. (**C**) Pyramidal neurons showed a significant difference in mean soma length (***$p = 6.3 \times 10^{-8}$, two-sided Welch's *t*-test, Holm–Bonferroni corrected). (**D**) Interneurons showed no significant difference in mean soma length.

(E)

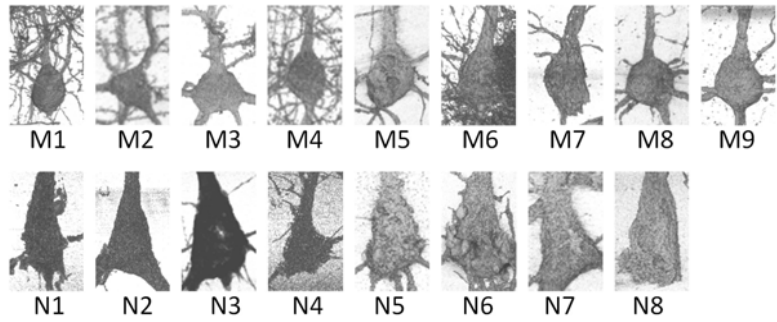

**Figure 2** (continued). (**E**) Renderings of pyramidal soma. Mouse individuals (M1-M9) and human cases (N1-N8) are indicated with labels. Human renderings are reproduced from Mizutani et al., 2019, 2023. Image width × height: 25 μm × 40 μm.

Structural differences in the three-dimensional neuronal network were also examined. Typical neurite structures are shown in Figure 3A and 3B. Major differences appear in the thickness and the tortuousness of the neurites. The tortuousness property can be represented with a parameter called curvature, which is reciprocal to the radius of the spatial trajectory. Mean neurite curvature for mice is over 1.8 times as large as that of humans (Figure 3C; $p$ = 0.0040). The neurite thickness of mice is less than 60% of that of humans (Figure 3D; $p$ = 0.0111). These results indicate that mouse neurites are significantly thinner and more tortuous compared with human neurites. The neurite thickness and curvature showed a reciprocal correlation (Figure 3E), which was also observed for human neurons (Mizutani et al., 2019, 2021).

The differences between human and mouse dendritic spine were rather moderate (Figure 3F-I). The spine curvature for mice was approximately 25% larger than that of the human controls (Figure 3F; $p$ = 0.0063), but no significant differences were found for the mean diameter (Figure 3G) or for the length (Figure 3H). The mean spine density for mice was 2.6 times as high as that of humans (Figure 3I; $p$ = 0.0070), though the staining efficiency of the Golgi method used in this study should have affected the density estimates.

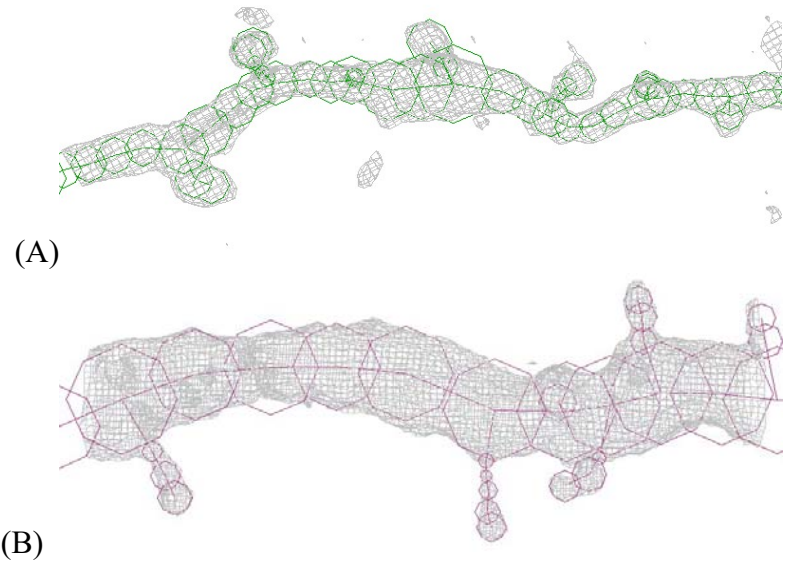

(A)

(B)

**Figure 3.** Differences in neurite and spine. (**A**) Mouse neurite indicated with the red box in Figure 1B. Traced structures (green) are superposed on the observed image (gray). The image is contoured at three times the standard deviation (3 σ) of the voxel intensity with a grid spacing of 96.2 nm. (**B**) Human neurite indicated with the red box in Figure 1D. The image is contoured at 3 σ with a grid of 97.4 nm.

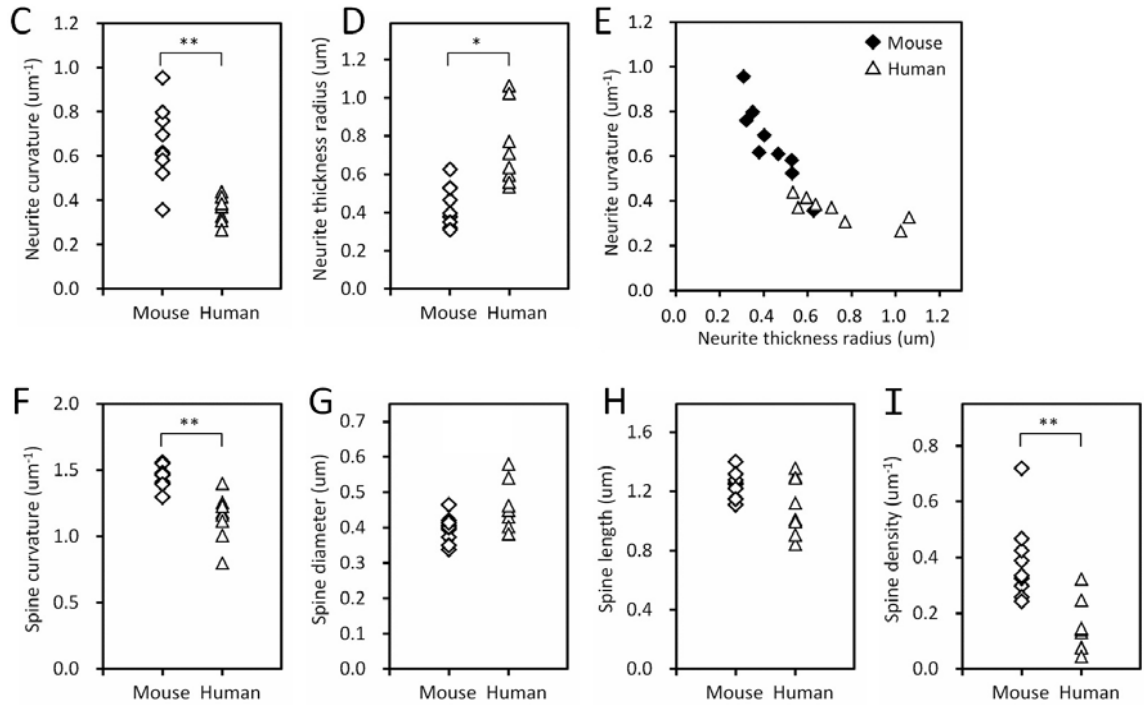

**Figure 3 (cont'd).** Differences in neurite and spine. (**C**) Difference in neurite curvature (**\*\****p* = 0.0040, two-sided Welch's *t*-test, Holm–Bonferroni corrected, same hereafter). (**D**) Neurite thickness radius (\**p* = 0.0111). (**E**) Scatter plot of neurite curvature and thickness illustrates their reciprocal correlation. Mouse data are indicated with diamonds and human data with triangles. (**F**) Spine curvature (**\*\****p* = 0.0063). (**G**) Spine diameter and (**H**) spine length showed no significant difference. (**I**) Spine density (**\*\****p* = 0.0070).

**Implementation of structural differences in generative AIs**

    Major structural differences between human and mouse were found in 1) the shape of the pyramidal soma and in 2) the thickness and curvature of the neurites. These differences should be related to each other, and this issue is addressed in the discussion section below. The thin and tortuous neurites in the mouse brain should suppress connections between distant neurons according to cable theory (Hansson et al., 1994).

We incorporated these findings in the generator of a deep convolutional generative adversarial network (DC-GAN; Radford et al., 2016) and examined its performance in image generation tasks. We recently reported a schizophrenia-mimicking convolutional layer, in which inter-node connections are suppressed depending on a distance between nodes (Mizutani et al., 2022). In this study, the inter-node distance was defined on the basis of two-dimensional arrangements of nodes (Figure 4A) to reproduce the laminar organization of neurons in the cerebral cortex. Connections outside the window cone in Figure 4A were kept at zero in the weight matrix in order to confine the connections to be within the cone.

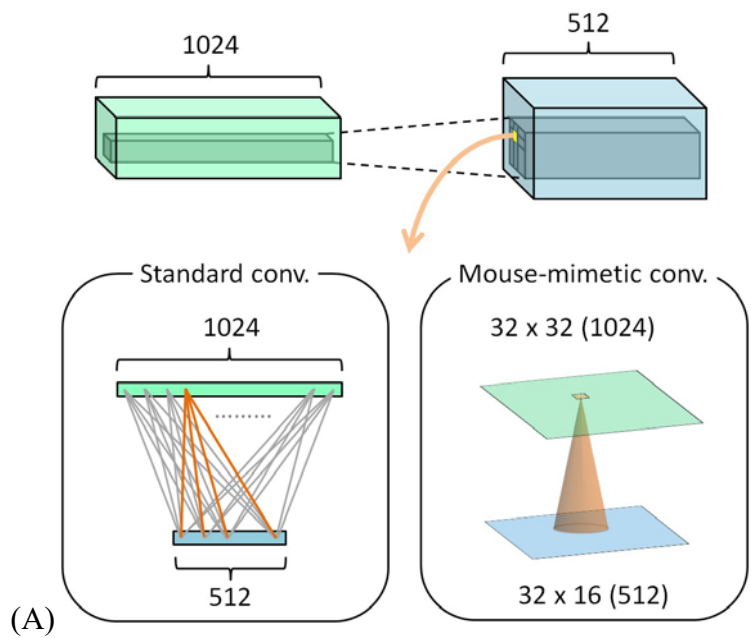

(A)

**Figure 4.** Image generation using mouse-mimetic AIs. (**A**) Schematic representation of convolutional layer. The standard convolutional layer is fully connected along the channel dimensions, while the mouse-mimetic layer is partially connected. Nodes in the mouse-mimetic layer are arranged in a two-dimensional manner to reproduce the laminar organization of neurons. An example of the connection window is shown with a cone. Weights within the cone are used in the training and evaluation, while those outside the cone are kept at zero. This results in a reduction of parameter usage in the mouse-mimetic layer. The degree of connection is defined by the radius of the cone base circle, which is expressed as a fraction of the total width of the node plane.

This mouse-mimetic convolutional layer was used as the hidden layers of the DC-GAN generator (Supplementary Table S5A), while other configurations of the model were kept the same as in the original report (Radford et al., 2016). The reduction in the number of weights in the generator is summarized in Supplementary Table S6. The GAN discriminator is composed of four standard convolutional layers and a fully-connected top layer (Supplementary Table S5B). GANs consisting of these generators and discriminators were trained from scratch by using photo datasets of cat faces (Animal Faces HQ; Choi et al., 2020), cheese (Cheese Pics), human faces (CelebA; Liu et al., 2015), or birds (Birds 525 Species). After the training, each GAN was used to generate fake photos, which were scored with the Fréchet inception distance (FID; Heusel et al., 2017).

The obtained results indicated that the mouse-mimetic GAN outperformed the standard GAN in the image generation task using the cat faces and cheese datasets, but underperformed for human faces and birds (Figure 4B, 4C, Supplementary Table S7A). Examples of generated photos (Supplementary Figure S21) of cat faces showed a slight mode collapse without using mouse-mimetic layers. Indeed, the FID scores for cat faces and cheese decreased as the parameter usage in the mouse-mimetic layers was reduced (Figure 4B). Linear regression of the FID scores to the parameter usage ratio showed positive slopes ($\beta = 0.121$, $p = 0.026$ for cat faces between 20–100%; $\beta = 0.38$, $p = 1.81 \times 10^{-6}$ for cheese between 35–100%, Holm–Bonferroni corrected). These results indicate that the mouse-mimetic GAN outperformed the standard GAN on the cat faces and cheese datasets. The best FID score for cat faces was obtained when only 35% of the weights were used in the mouse-mimetic layers. Whereas the standard GAN failed to converge in seven out of ten training runs using the cat faces dataset (Supplementary Table S7A), the mouse-mimetic GAN with a parameter usage of 35% or less failed to converge in only one out of 20 training runs. The course of the FID scores (Supplementary Figure S22) indicated that the incorporation of the mouse-mimetic layers stabilized the GAN. We also implemented mouse-mimetic layers in the discriminator (Supplementary Table S5B) and trained it with the cat faces dataset, though all ten training runs failed to converge (Supplementary Table S7A), indicating that the mouse-mimetic design is effective only in the generator. Contrary to the results of the cat faces and cheese datasets, the FID scores for human faces and birds increased as the parameter usage was reduced (Figure 4C; linear regression between 35–100%: $\beta = -0.164$, $p = 3.4 \times 10^{-9}$ for human faces; $\beta = -0.21$, $p = 9.2 \times 10^{-6}$ for birds, Holm–Bonferroni corrected). These results indicate that the mouse-mimetic GAN underperformed the standard GAN in generating images of human faces and birds.

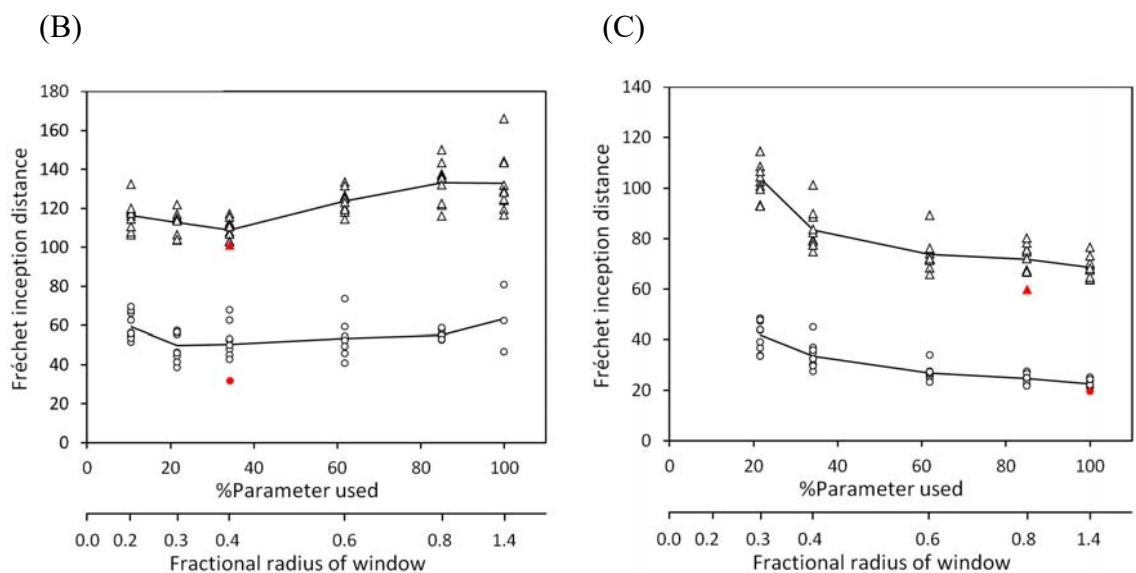

**Figure 4 (cont'd).** Image generation using mouse-mimetic AIs. (**B**) Fréchet inception distance (FID) scores of cat faces (circles) and cheese (triangles) photos generated using a generative adversarial network (GAN) are plotted against the percent usage of weights in the mouse-mimetic layers and against the fractional radius of the window. 100% use of parameters corresponds to the standard network. The training and evaluation were repeated for ten runs for each %usage. A total of 12 runs using the cat faces dataset did not converge and had FID scores greater than 200 (Supplementary Table S7A). Lines indicate the mean FID scores of the converged runs. Red symbols indicate the best FID scores. (**C**) FID scores of human faces (circles) and bird (triangles) photos generated using the GAN. All runs converged, and their FID scores are plotted. Lines indicate mean FID scores. Red symbols indicate the best FID scores.

Next, we examined a conditional GAN (Mirza & Osindero, 2014) model by using a merged dataset composed of the above four datasets. Approximately 5,000 images from each dataset were extracted by sharding and used for training the conditional GAN, of which the generator and discriminator were the same as those in Supplementary Table 5, except for adding image class inputs. The parameter usage was set to 34.2% in the mouse-mimetic layers. The FID scores of the photos generated for the human-face and bird image classes increased as a result of incorporating mouse-mimetic layers (Supplementary Figure S23; Supplementary Table S7B; human face: $p = 1.10 \times 10^{-5}$; bird: $p = 0.0114$). This indicates that the mouse-mimetic GAN underperformed the standard GAN. In contrast, the FID scores for cat faces and cheese showed larger variances and no significant difference between the mouse-mimetic and standard GANs (Supplementary Figure S23; Supplementary Table S7B). The best FID score for cheese was obtained using the mouse-mimetic conditional GAN.

We also implemented mouse-mimetic convolutional layers in the U-Net (Ronneberger et al., 2015) of a denoising diffusion implicit model (DDIM; Song et al., 2021) and trained the network on the above four datasets individually. The mouse-mimetic DDIM, in which only 44% of the weights of the U-Net were used (Supplementary Table S8), outperformed the control standard DDIM in the image generation task using the cat faces and the cheese datasets (cat faces: $p = 0.000131$; cheese: $p = 0.00088$; Figure 4D, Supplementary Table S7C). In contrast, the mouse-mimetic DDIM underperformed the standard DDIM for birds ($p = 0.00089$) and showed no significant difference for human faces (Figure 4D, Supplementary Table S7C). Examples of the generated images are shown in Figure 4E and Supplementary Figure S25. The cat photo examples generated by the standard DDIM seem dominated by tabby cats, while those of the mouse DDIM seem rather diverse. The mouse-mimetic DDIM was slower to converge on the FID score, but was more stable and less overfitting than the standard DDIM (Supplementary Figure S24). These results indicate that 1) the number of parameters of the U-Net in the DDIM can be reduced to less than half the original U-Net by introducing mouse-mimetic layers and that 2) the mouse-mimetic parameter reduction can improve the DDIM performance under certain conditions.

(D)

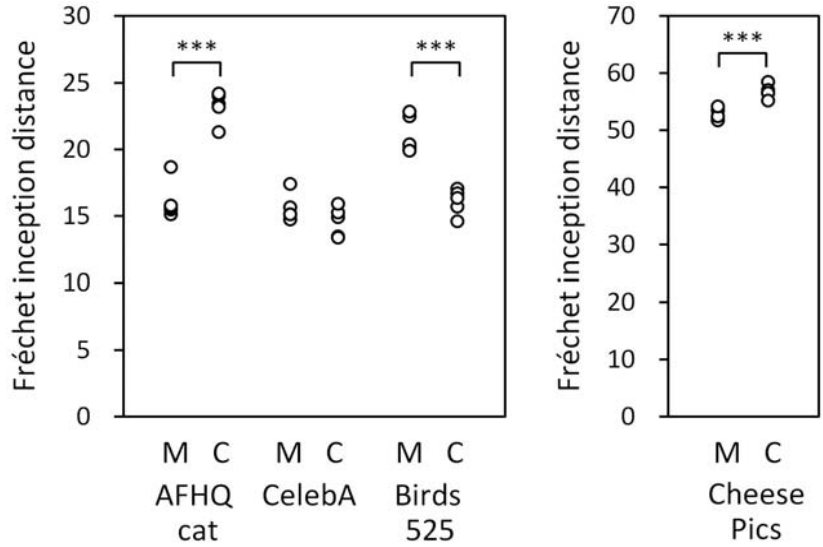

**Figure 4 (cont'd).** Image generation using mouse-mimetic AIs. (**D**) FID scores of photos generated using denoising diffusion implicit models (DDIMs). The parameter usage was set to 44% in the mouse-mimetic DDIM (marked 'M') and 100% in the control standard DDIM (marked 'C'). Labels indicate the datasets used for training (AFHQ cat: cat faces; CelebA: human faces; Birds 525: birds; Cheese Pics: cheese). Differences in the FID scores between the mouse-mimetic and standard DDIMs were examined using a two-sided Welch's *t*-test and corrected with the Holm–Bonferroni method (AFHQ cat: ***$p$ = 0.000131; Birds 525 Species: ***$p$ = 0.00089; Cheese Pics: ***$p$ = 0.00088).

(E)

Mouse-mimetic

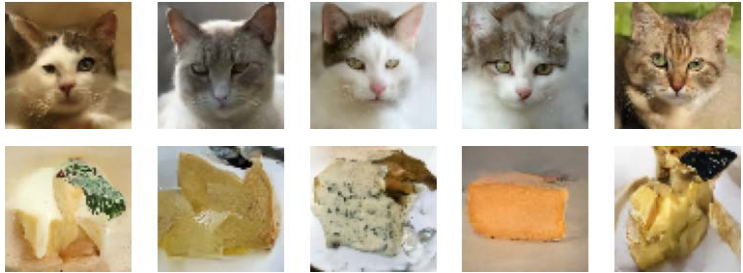

Standard

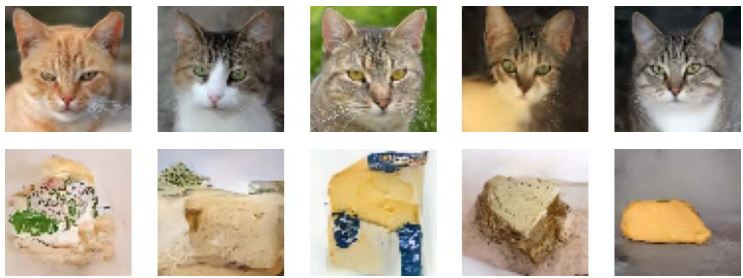

**Figure 4 (cont'd).** Image generation using mouse-mimetic AIs. (**E**) Example images of cat faces and cheese generated from the best runs of the mouse-mimetic and standard DDIMs.

Since the results of the GAN and the DDIM are similar independently of the network architecture and the image generation algorithm, the performance differences should be due to differences between datasets. We examined the statistics of the four datasets and found differences between their image entropy distributions (Figure 4F; Supplementary Table S9). Though the datasets are nearly the same in terms of the mean image entropy, the standard deviation of the image entropy differs between the datasets. The frequency distribution of the image entropy of the Animal Faces HQ cat faces dataset showed a sharp peak (Supplementary Figure S26A), indicating that the images of this dataset have similar intensity histograms. In contrast, the Cheese Pics dataset showed tails toward the low entropy side (Supplementary Figure S26B). This represents that the dataset contains images showing uniform intensity, such as those of cheese blocks. The CelebA human faces and Birds 525 Species datasets showed profiles in between them (Supplementary Figure S26C, D). The results of an AKAZE local feature matching (Alcantarilla et al., 2011) indicated that the cheese dataset is different from the other three datasets in terms of the standard deviation of the matching distance (Figure 4F). The number of detected feature points differed between the datasets, though the cheese and bird datasets showed similar values (Figure 4F). Another difference between the datasets was in the number of images constituting each dataset (Supplementary Table S9). We prepared a partial dataset consisting of 5199 images from the CelebA dataset and trained the above unconditional GAN on it. The obtained FID scores did not show a positive slope, and the best FID was obtained by using the standard GAN, as in the case of the complete CelebA dataset (Supplementary Figure S27). These results indicate that the dataset size itself does not determine the relationship between the FID score and the parameter usage ratio.

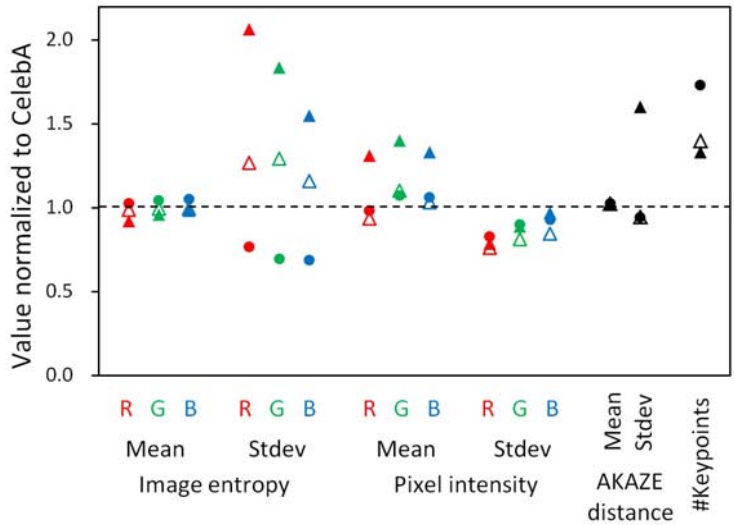

**Figure 4 (cont'd).** Image generation using mouse-mimetic AIs. (**F**) Statistics of datasets used for training. Values are normalized to those of the CelebA human faces dataset. Circles indicate statistics for the Animal Faces HQ cat dataset, closed triangles those for Cheese Pics, and open triangles Birds 525 Species. Color channels are color-coded.

**Discussion**

    The results of this study indicated that the neurons of human and mouse differ in a number of structural parameters, including soma length (Figure 2) and neurite thickness (Figure 3D). A difference in size of neuronal soma between humans and mice was also reported for hippocampal CA1 neurons (Benavides-Piccione et al., 2020). The soma downsizing and the neurite thinning allow neurons to be integrated in a limited thickness and area of the mouse cerebral cortex and, therefore, should be the results of adaptation to the limited volume of the mouse brain. In contrast to the mouse brain, the human brain occupies a large volume and bears a thick and folded cerebral cortex (Van Essen & Drury, 1997; Im et al., 2008; Kang et al., 2012; Schnack et al., 2015), which can accommodate long and thick neurons. If mouse neurons are integrated into the volume of the human brain, the number of neurons per cortical area can be increased by two times or more and the amount of information that are processed simultaneously can be doubled, though the energy supply issue (Roy & Sherrington, 1890; Leybaert, 2005; Saiga et al., 2021) remains.

Since the mean stem-dendrite diameter is proportional to the neuronal soma diameter (Zwaagstra & Kernell, 1981), the differences in soma size and neurite diameter should be related to each other. The soma size correlates also with the dendrite extent (Fiala & Harris, 1999). These results along with ours suggest that a neuron with a large soma has thick dendrites innervating widely, while a neuron with a small soma has thin dendrites wherein the network is confined to the vicinity of the soma. It has been reported that a neuregulin-4 knock-out mouse showed a reduced soma size of pyramidal neurons in the motor cortex and exhibited defects in motor performance (Paramo et al., 2021). A mouse model of Rett syndrome exhibited reduced dendritic arborization (Jentarra et al., 2010) and a reduced soma size (Rangasamy et al., 2016). Therefore, the soma downsizing and the neurite thinning in the mouse brain should correlate with each other and affect the performance of the neuronal network.

Differences between the dendritic spines were observed in their curvature and density (Figure 3F-I). Since the spine curvature reciprocally correlates with the spine thickness (Mizutani et al., 2021) and the spine thickness correlates with the neurite thickness (Mizutani et al., 2021), the tortuous spines in the mouse brain should affect the neuronal network concomitantly with the neurite thinning. The mouse spine density determined with our method was 2.6 times as high as that of humans (Figure 3I). Though the staining efficiency of the Golgi impregnation used in our studies should have affected the density estimation, a similar difference in synapse density between humans and mice was reported in a dense connectomic study (Loomba et al., 2022). The two- to three-fold higher spine density in mice can compensate for the 2- to 3-fold thinner cortex compared to the human cortex, allowing neurons to form a comparable number of spines per cortical area. We suggest that dendritic spines of humans and mice are regulated to keep their areal density constant.

This study compared three-dimensional structures of neurons between the medial prefrontal cortex of mice and the anterior cingulate cortex of humans. The comparison revealed multiple differences as described above. It has been reported that the anterior cingulate cortex of humans exerts emotional and cognitive functions (Botvinick et al., 1999; Bush et al., 2000). Moreover, neurons in the medial prefrontal cortex of mice receive long-range inputs from brain regions involved in cognition, motivation, and emotion (Anastasiades & Carter, 2022). Therefore, the neuronal differences observed in this study should be relevant to cognitive and emotional brain functions in humans and mice, including functions for finding food and fleeing from enemies.

The structural comparison of human and mouse neurons suggested that neuronal connections are spatially confined in the mouse brain network. We incorporated this finding in DC-GAN and DDIM by masking the weight matrix depending on the two-dimensional distance between nodes. The resultant FID scores for human faces and birds were worsened by the parameter reduction for both the GAN and the DDIM (Figure 4C, 4D). The constraints in the weight matrix should have degraded the performance of the generative AIs. In contrast, the FID scores for cat faces and cheese were improved by the parameter reduction (Figure 4B, 4D). The sparse weight matrix of the mouse-mimetic layer can eliminate extra degrees of freedom and thus may have improved the performance. These results suggest that the number of parameters appropriate for generating cat faces and cheese photos is lower than those for human faces and birds. The four datasets used in this study showed differences in image entropy, which can affect the number of parameters required for the image generation task. We suggest that the statistics of the input dataset including the image entropy determine the degree of freedom suited for generating images from that dataset.

The results for the mouse-mimetic networks indicated that the number of parameters can be adjusted by using the mouse-mimetic layers without changing the network architecture or dimensions. Since the shape of the weight matrix of the mouse layer is the same with that of its original one, any convolutional layer can be replaced with its mouse-mimetic version. Although the parameter %usage needs to be optimized to obtain the best result, we suggest 30–50% as a first choice to examine whether the mouse-mimetic layer works in the target application. Since it has been reported that the brain-wide circuitry associated with higher intelligence is organized in a sparse manner (Genç et al., 2018), parameter reduction should also be a strategy to improve the performance in biological systems. The results of this study indicated that the number of parameters used for the image generation task can be halved by introducing mouse-mimetic layers. Although the computation time for training the mouse-mimetic AI is the same with that of its standard version in the present implementation, there is a possibility to reduce the computational cost by taking account of the sparse architecture of the mouse-mimetic layer into the library code of the neural network calculation.

The ages of the mice used in this study were in the young adult range, while the human cases ranged in age from their early 40's to 70's (Mizutani et al, 2023). This age difference can affect the structure of the neuronal network, though a significant correlation between the age and the structural parameter was observed only for the standard deviation of the neurite curvature in the human study (Mizutani et al., 2023).

We suggest that age-related effects on the human neuronal structure are rather limited compared with the differences between human and mouse. Another limitation of this study is the difference in the tissue fixation method. The mouse tissues were fixed with a perfusion fixation, while the human tissues were fixed by immersing them in a fixation solution. This methodological difference can affect neuron structures.

In this study, we analyzed three-dimensional structures of mouse brain tissues and compared them with those of human brain tissues. The neurons in the mouse brain showed comparably downsized somata and thin neurites, allowing the neurons to be integrated in a limited space of the mouse brain. This finding was applied to generative AIs to examine its computational consequences. The mouse-mimetic AIs outperformed the conventional AIs in image generation tasks using cat faces and cheese datasets. The preferences of the mouse-mimetic AIs coincided with the impressions commonly associated with mice, though its biological implication remains to be clarified. The structure of mouse neurons would have adapted to the environment that mice live in to optimize their brain functions for their survival. We suggest that relationship between neuronal structure and brain function should be investigated by implementing other biological findings in artificial neural networks.

## Methods

### Human cerebral data

Human data were obtained in our previous studies (Mizutani et al., 2019, 2023). Post-mortem human cerebral tissues were collected with written informed consent from the legal next of kin using protocols approved by the Clinical Study Reviewing Board of Tokai University School of Medicine (application no. 07R-018). The human studies were conducted according to the Declaration of Helsinki. The method used for the structural analysis of human tissues (Mizutani et al., 2019, 2023) is the same as the method applied to the mouse samples described below.

### Mouse cerebral tissue

Nine male C57BL/6J mice were housed in a conventional environment with ad libitum feeding under a 12 h / 12 h light and dark cycle. All mouse experiments were performed with the approval from the Institutional Animal Care and Use Committee at the Tokyo Metropolitan Institute of Medical Science (protocol code: 18028) in accordance with relevant guidelines and regulations. The mice at 16 weeks of age were deeply anesthetized with an overdose of sodium pentobarbital and subjected to perfusion fixation using phosphate-buffered saline containing 4% formaldehyde. Brains were dissected and stained with Golgi impregnation. Left medial prefrontal cortexes were taken from the stained tissues and embedded in epoxy resin. The staining and embedding procedures were performed as reported previously (Mizutani et al., 2019).

**Synchrotron radiation X-ray microtomography and nanotomography**

The resin-embedded samples were first subjected to simple projection microtomography at the BL20XU beamline (Suzuki et al., 2004) of the SPring-8 synchrotron radiation facility to visualize the structure of the entire tissue. The data collection conditions are summarized in Supplementary Table S1. Absorption contrast X-ray images were collected using a visible-light conversion type X-ray imaging detector consisting of a phosphor screen, optical lenses and complementary metal-oxide semiconductor (CMOS) camera (ORCA-Flash4.0, Hamamatsu Photonics, Japan), as reported previously (Mizutani et al., 2019). Three-dimensional structures of the samples were reconstructed with the convolution back-projection method in order to determine layer positions. An example of the structure is shown in Supplementary Figure S28.

Tissue volumes corresponding to layer V were then visualized with synchrotron radiation X-ray nanotomography (nano-CT) equipped with Fresnel zone plate optics (Takeuchi et al., 2002). The nano-CT experiments were performed as reported previously (Mizutani et al., 2019, 2021, 2023) at the BL37XU (Suzuki et al., 2016) and the BL47XU (Takeuchi et al., 2009) beamlines of SPring-8, and at the 32-ID beamline (De Andrade et al., 2016, 2021) of the Advanced Photon Source of Argonne National Laboratory. The data collection conditions are summarized in Supplementary Table S1. Photon flux at the sample position in the BL47XU experiment was determined to be $1.0 \times 10^{15}$ photons/mm$^2$/s by using Al$_2$O$_3$:C dosimeters (Nagase-Landauer, Japan). Photon fluxes at the BL37XU and at the 32-ID beamlines were reported previously (Mizutani et al., 2023). Spatial resolutions were estimated from the Fourier domain plot (Mizutani et al., 2016) or by using three-dimensional test patterns (Mizutani et al., 2008).

**Tissue structure analysis**

The obtained datasets were processed with the convolution-back-projection method in the RecView software (RRID: SCR_016531; https://mizutanilab.github.io/) (Mizutani et al., 2010) to reconstruct three-dimensional images (Figure 1A), as reported previously (Mizutani et al., 2019, 2023). The image reconstruction was conducted by RS and KS. The obtained image datasets were shuffled with human datasets from our previous study (Mizutani et al., 2023) and provided to RM without any data attributes. RM built Cartesian coordinate models of tissue structures from the three-dimensional images using the MCTrace software (RRID: SCR_016532; https://mizutanilab.github.io/) (Mizutani et al., 2013). After the model building of each dataset finished, the coordinate files of the structural models were locked down. Then, RM reported the number of neurite segments to RS. RS aggregated the numbers in order to equalize analysis amounts between mouse individuals and between human and mouse. Datasets that should be further analyzed were determined by RS and provided to RM without any data attributes or aggregation results.

These model building procedures were repeated for three batches of nano-CT datasets. The first two batches consisted of eight mouse datasets (M1A, M1B, M1C, M2A, M2B, M3A, M3B, and M4A) and 12 human datasets of our previous study (Mizutani et al., 2023). The information of these datasets was disclosed to RM after the model building of these two batches was finished in order to report the human study (Mizutani et al., 2023). The other batch consisted of 13 mouse datasets, two dummy mouse datasets unrelated to this study, and eight human datasets. The dummy mouse and human datasets were included in order to shuffle the datasets. After the model building of the entire batch finished, the dataset collection date was first disclosed to RM to correct the voxel width which was tentatively defined. After the voxel width was corrected, the coordinate files were locked down and all attributes of the datasets were opened to assign each dataset to mouse individuals.

The human datasets and the two dummy mouse datasets unrelated to this study were not used in the subsequent analysis. All other 21 datasets were subjected to structural analysis. Structural parameters were calculated from Cartesian coordinate models by using the MCTrace software, as reported previously (Mizutani et al., 2019, 2023). The soma width was defined with the diameter of the largest spherical nodes composing the neuron. The soma length was defined as the length along spherical nodes having a diameter larger than half the soma width (Figure 2A). Statistics of the obtained structural parameters are summarized in Supplementary Tables S2–S4.

**Photo image datasets**

The photo image datasets used for training generative AIs were taken from hyperlinks in the Datasets page of the kaggle.com website (https://www.kaggle.com/datasets). The CelebA dataset (Liu et al., 2015) was used as the dataset of human-face photos. A partial CelebA dataset consisting of the first 5199 images in numerical filename order was also prepared to examine the effect of the number of images per dataset. The train-cat folder of the Animal Faces HQ dataset (Choi et al., 2020) was used as the dataset of cat face photos. The Birds 525 Species dataset was used as the dataset of bird photos. The Cheese Pics dataset was used as the source of cheese photos. Since the Cheese Pics dataset contains photos of humans, buildings, packages, and cooked foods, we examined the contents of all 1824 folders of this dataset and chose 165 folders (Supplementary Table S10) which mainly contain cheese photos. All images of these datasets were scaled to 64 × 64 pixel dimensions and to pixel values in a 0–1 range prior to training.

The analysis of the dataset statistics and the AKAZE local feature matching (Alcantarilla et al., 2011) were conducted using the OpenCV-4.10.0 library (https://opencv.org/releases/). The AKAZE feature matching was performed using the first 2000 images of each dataset. Prior to the feature point detection, the images in the 64 × 64 pixel dimensions were converted to gray scale and resized to 256 × 256 pixels.

**Generative adversarial network**

The structural analysis of mouse cerebral tissue indicated that mouse neurites are thin and tortuous compared with those of humans. This means that neuronal connections are confined depending on the distance between neurons. We incorporated this finding in an artificial neural network by masking weights with a window matrix, which are multiplied with the weight matrix in an element-by-element manner (Mizutani et al., 2022). In this study, the inter-node distance was defined by assuming a two-dimensional arrangement of nodes (node position $x$, $y$) in order to mimic the laminar organization of neurons in the cerebral cortex. The inter-node distance was defined along the channel dimensions (Figure 4A). Fractional coordinates ($x$ / total nodes along $x$ dimension, $y$ / total nodes along $y$ dimension) were used to calculate the Euclidean distance between nodes. Elements of the window matrix were set to 1 if the fractional distance between a node pair is less than a predefined threshold, and set to 0 if the distance is equal to or larger than the threshold.

The mouse-mimetic convolutional layers with the two-dimensional window were implemented in the generator of the DC-GAN (Supplementary Table S5A; Radford et al., 2016). The discriminator of the GAN was made of four standard convolutional layers and a fully-connected top layer (Supplementary Table S5B). Batch normalization (Ioffe & Szegedy, 2015) was applied to hidden layers of the generator, and spectral normalization (Miyato et al., 2018) to hidden layers of the discriminator. ReLU activation function was used in the hidden layers of the generator, and leaky ReLU in the hidden layers of the discriminator.

Conditional GAN (Mirza & Osindero, 2014) was performed by using the same configurations as the unconditional GAN described above, except for adding image class inputs. The datasets of CelebA, Birds 525 Species, and Cheese Pics were sharded in order to extract approximately 5,000 images for each image class. The extracted images of the three image classes along with the entire Animal Faces HQ cat faces dataset were merged and shuffled to compose a class-labeled dataset used for training the conditional GAN model.

These GAN models were trained from scratch using the Adam algorithm (Kingma & Ba, 2015) with a batch size of 32, learning rates of $1 \times 10^{-4}$, and $\beta_1$ of 0.5 for both the generator and the discriminator. The discriminator was trained once per cycle. The number of training epochs was set to 25 for the CelebA dataset. This corresponds to approximately 160,000 steps of training. The number of training epochs for other datasets was determined so as to set the total number of steps to be approximately 80,000. The Fréchet inception distance (FID; Heusel et al., 2017) was calculated from 51200 images by using the code provided at https://github.com/jleinonen/keras-fid. The training and evaluation were repeated for 10 runs. Examples of the FID score progress during training are shown in Supplementary Figure S22. The calculations were performed using Tensorflow 2.16.1 and Keras 3.3.3 running on the g4dn.xlarge instance (NVIDIA T4 Tensor Core GPU with Intel Xeon Cascade Lake P-8259CL processor operated at 2.5 GHz) of Amazon Web Service. The training for 80,000 steps took typically 3 hours. Python codes are available from our GitHub repository (https://mizutanilab.github.io).

**Denoising diffusion implicit model**

The mouse-mimetic convolutional layers were implemented in the U-Net (Ronneberger et al., 2015) of the DDIM (Song et al., 2021). The U-Net was downsized to [64, 128, 256, 512] channels from [128, 256, 256, 256, 512] channels of the original implementation for the CelebA dataset (Song et al., 2021) to halve the computation time required for training. The mouse-mimetic parameter reduction was applied to the convolutional layers of the residual blocks with channel dimensions of 256 and 512. The parameter %usage is summarized in Supplementary Table S8. Attention layers were introduced at the 16 × 16 feature map resolution. Group normalization (Wu & He, 2018) and a swish activation function (Ramachandran et al., 2017) were used in accordance with the GitHub repository of the original DDIM report (Song et al., 2021).

The DDIM was trained from scratch using the Adam algorithm (Kingma & Ba, 2014) with a batch size of 64, learning rate of $2 \times 10^{-4}$, and $\beta_1$ of 0.9. The number of training epochs was determined so as to set the total number of steps to be approximately 500,000. The Fréchet inception distance (FID; Heusel et al., 2017) was calculated from 51200 images generated by 20 diffusion steps. The training and evaluation were repeated for 5 runs using the same environment as the GAN. Examples of the FID scores' progress during training are shown in Supplementary Figure S24. The DDIM training for 500,000 steps typically took about 70 hours. Python codes are available from our GitHub repository (https://mizutanilab.github.io).

**Statistical tests**

The statistical tests of the structural parameters and FID scores were performed using the R software, as reported previously (Mizutani et al., 2019, 2021, 2022, 2023). Significance was defined as $p < 0.05$. Differences in the means of the structural parameters between human and mouse were examined using a two-sided Welch's $t$-test. The relation between the mean FID score and the mean parameter usage ratio was examined by linear regression analysis. Differences in the means of the FID scores were examined using a two-sided Welch's $t$-test. The $p$-values of multiple tests were corrected with the Holm–Bonferroni method.

**Acknowledgements**
We thank the Technical Service Coordination Office of Tokai University for assistance in preparing sample adapters for nano-CT. The studies on human neurons were conducted under the approval of the Clinical Study Reviewing Board of Tokai University School of Medicine and the Institutional Biosafety Committee of Argonne National Laboratory. Mouse experiments were performed with the approval of the Institutional Animal Care and Use Committee of the Tokyo Metropolitan Institute of Medical Science. This research was supported by the Japan Society for the Promotion of Science KAKENHI (23H02800 and 23K27491 to RM; 22H04922 to AdAMS). The synchrotron radiation experiments at SPring-8 were performed with the approval of the Japan Synchrotron Radiation Research Institute (JASRI) (proposal nos. 2019B1087, 2021A1175, 2021B1258, and 2023B1187). The synchrotron radiation experiment at the Advanced Photon Source of Argonne National Laboratory was performed under General User Proposals GUP-59766 and GUP-78336. This research used resources of the Advanced Photon Source, a U.S. Department of Energy (DOE) Office of Science User Facility operated for the DOE Office of Science by Argonne National Laboratory under Contract No. DE-AC02-06CH11357.

**Conflict of interest**

All authors declare no conflict of interest.

**References**

1. D C Van Essen, H A Drury (1997). Structural and Functional Analyses of Human Cerebral Cortex Using a Surface-Based Atlas. J Neurosci 17(18): 7079–7102.
2. Kiho Im, Jong-Min Lee, Oliver Lyttelton, Sun Hyung Kim, Alan C Evans, Sun I Kim (2008). Brain size and cortical structure in the adult human brain. Cereb Cortex 18(9): 2181–2191.
3. Xiaojian Kang, Timothy J Herron, Anthony D Cate, E William Yund, David L Woods (2012). Hemispherically-unified surface maps of human cerebral cortex: reliability and hemispheric asymmetries. PLoS One 7(9): e45582.
4. Hugo G Schnack, Neeltje E M van Haren, Rachel M Brouwer, Alan Evans, Sarah Durston, Dorret I Boomsma, René S Kahn, Hilleke E Hulshoff Pol (2015). Changes in thickness and surface area of the human cortex and their relationship with intelligence. Cereb Cortex 25(6): 1608–1617.
5. B Fischl, A M Dale (2000). Measuring the thickness of the human cerebral cortex from magnetic resonance images. Proc Natl Acad Sci U S A 97(20): 11050–11055.
6. Konrad Wagstyl et al. (2020). BigBrain 3D atlas of cortical layers: Cortical and laminar thickness gradients diverge in sensory and motor cortices. PLoS Biol 18(4): e3000678.
7. C C Larsen, K Bonde Larsen, N Bogdanovic, H Laursen, N Graem, G Badsberg Samuelsen, B Pakkenberg (2006). Total number of cells in the human newborn telencephalic wall. Neuroscience 139(3): 999–1003.
8. Suzana Herculano-Houzel (2009). The human brain in numbers: a linearly scaled-up primate brain. Front Hum Neurosci 3: 31.
9. Christopher S von Bartheld, Jami Bahney, Suzana Herculano-Houzel (2016). The search for true numbers of neurons and glial cells in the human brain: A review of 150 years of cell counting. J Comp Neurol 524(18): 3865–3895.
10. Marco Pagani, Mario Damiano, Alberto Galbusera, Sotirios A Tsaftaris, Alessandro Gozzi (2016). Semi-automated registration-based anatomical labelling, voxel based morphometry and cortical thickness mapping of the mouse brain. J Neurosci Methods 267: 62–73.
11. Luam Hammelrath, Siniša Škokić, Artem Khmelinskii, Andreas Hess, Noortje van der Knaap, Marius Staring, Boudewijn P F Lelieveldt, Dirk Wiedermann, Mathias Hoehn (2016). Morphological maturation of the mouse brain: An in vivo MRI and histology investigation. Neuroimage 125: 144–152.

12. Daniel Keller, Csaba Erö, Henry Markram (2018). Cell densities in the mouse brain: A systematic review. Front Neuroanat 12: 83.
13. Laurence H Tecott (2003). The genes and brains of mice and men. Am J Psychiatry 160(4): 646–656.
14. Inmaculada Ballesteros Yáñez, Alberto Muñoz, Julio Contreras, Juncal Gonzalez, Elisia Rodriguez-Veiga, Javier DeFelipe (2005). Double bouquet cell in the human cerebral cortex and a comparison with other mammals. J Comp Neurol 486: 344–360.
15. Andrew D Strand, Aaron K Aragaki, Zachary C Baquet, Angela Hodges, Philip Cunningham, Peter Holmans, Kevin R Jones, Lesley Jones, Charles Kooperberg, James M Olson (2007). Conservation of regional gene expression in mouse and human brain. PLoS Genet 3(4): e59.
16. Steffen Katzner, Sarah Weigelt (2013). Visual cortical networks: of mice and men. Curr Opin Neurobiol 23(2): 202–206.
17. Rebecca D Hodge, Trygve E Bakken, Jeremy A Miller, Kimberly A Smith, Eliza R Barkan, Lucas T Graybuck et al. (2019). Conserved cell types with divergent features in human versus mouse cortex. Nature 573: 61–68.
18. Ruth Benavides-Piccione, Mamen Regalado-Reyes, Isabel Fernaud-Espinosa, Asta Kastanauskaite, Silvia Tapia-González, Gonzalo León-Espinosa, Concepcion Rojo, Ricardo Insausti, Idan Segev, Javier DeFelipe (2020). Differential structure of hippocampal CA1 pyramidal neurons in the human and mouse. Cereb Cortex 30(2): 730–752.
19. Viktor Szegedi, Melinda Paizs, Judith Baka, Pál Barzó, Gábor Molnár, Gabor Tamas, Karri Lamsa (2020). Robust perisomatic GABAergic self-innervation inhibits basket cells in the human and mouse supragranular neocortex. eLife 9, e51691
20. Trygve E Bakken, Nikolas L Jorstad, Qiwen Hu, Blue B Lake, Wei Tian, Brian E Kalmbach et al. (2021). Comparative cellular analysis of motor cortex in human, marmoset and mouse. Nature 598(7879): 111–119.
21. Antoine Beauchamp, Yohan Yee, Ben C Darwin, Armin Raznahan, Rogier B Mars, Jason P Lerch (2022). Whole-brain comparison of rodent and human brains using spatial transcriptomics. eLife 11: e79418.
22. Sahil Loomba, Jakob Straehle, Vijayan Gangadharan, Natalie Heike, Abdelrahman Khalifa, Alessandro Motta, Niansheng Ju, Meike Sievers, Jens Gempt, Hanno S Meyer, Moritz Helmstaedter (2022). Connectomic comparison of mouse and human cortex. Science 377(6602): eabo0924.
23. Mean-Hwan Kim, Cristina Radaelli, Elliot R Thomsen, Deja Monet, Thomas Chartrand, Nikolas L Jorstad, et al. (2023). Target cell-specific synaptic dynamics of excitatory to inhibitory neuron connections in supragranular layers of human neocortex. eLife 12:

e81863.

24. Elisabeth Ragone, Jacob Tanner, Youngheun Jo, Farnaz Zamani Esfahlani, Joshua Faskowitz, Maria Pope, Ludovico Coletta, Alessandro Gozzi, Richard Betzel (2024). Modular subgraphs in large-scale connectomes underpin spontaneous co-fluctuation events in mouse and human brains. Commun Biol 7: 126.

25. R Mizutani, R Saiga, A Takeuchi, K Uesugi, Y Terada, Y Suzuki, V De Andrade, F De Carlo, S Takekoshi, C Inomoto, N Nakamura, I Kushima, S Iritani, N Ozaki, S Ide, K Ikeda, K Oshima, M Itokawa, M Arai (2019). Three-dimensional alteration of neurites in schizophrenia. Transl. Psychiatry 9: 85.

26. R Mizutani, R Saiga, Y Yamamoto, M Uesugi, A Takeuchi, K Uesugi, Y Terada, Y Suzuki, V De Andrade, F De Carlo, S Takekoshi, C Inomoto, N Nakamura, Y Torii, I Kushima, S Iritani, N Ozaki, K Oshima, M Itokawa, M Arai (2021). Structural diverseness of neurons between brain areas and between cases. Transl. Psychiatry 11: 49.

27. Ryuta Mizutani, Rino Saiga, Yoshiro Yamamoto, Masayuki Uesugi, Akihisa Takeuchi, Kentaro Uesugi, Yasuko Terada, Yoshio Suzuki, Vincent De Andrade, Francesco De Carlo, Susumu Takekoshi, Chie Inomoto, Naoya Nakamura, Youta Torii, Itaru Kushima, Shuji Iritani, Norio Ozaki, Kenichi Oshima, Masanari Itokawa, Makoto Arai (2023). Structural aging of human neurons is opposite of the changes in schizophrenia. PLoS One 18(6): e0287646.

28. Sara Scarfo, Antonella M A Marsella, Loulouda Grigoriadou, Yashar Moshfeghi, William J McGeown (2024). Neuroanatomical correlates and predictors of psychotic symptoms in Alzheimer's disease: a systematic review and meta-analysis. Neuropsychologia 24: 109006.

29. Yana R Panikratova, Irina S Lebedeva, Tatiana V Akhutina, Denis V Tikhonov, Vasilii G Kaleda, Roza M Vlasova (2023). Executive control of language in schizophrenia patients with history of auditory verbal hallucinations: A neuropsychological and resting-state fMRI study. Schizophr Res 262: 201–210.

30. Jingli Chen, Yarui Wei, Kangkang Xue, Shaoqiang Han, Wenbin Li, Bingqian Zhou, Jingliang Cheng (2024). Abnormal effective connectivity of reward network in first-episode schizophrenia with auditory verbal hallucinations. J Psychiatr Res 171: 207–214.

31. Demis Hassabis, Dharshan Kumaran, Christopher Summerfield, Matthew Botvinick (2017). Neuroscience-inspired artificial intelligence. Neuron 95(2): 245–258.

32. Tom Macpherson, Anne Churchland, Terry Sejnowski, James DiCarlo, Yukiyasu Kamitani, Hidehiko Takahashi, Takatoshi Hikida (2021). Natural and Artificial Intelligence: A brief introduction to the interplay between AI and neuroscience research. Neural Netw 144: 603–613.
30

33. K Fukushima (1980). Neocognitron: a self organizing neural network model for a mechanism of pattern recognition unaffected by shift in position. Biol Cybern 36(4): 193–202.
34. Toshiteru Homma, Les E Atlas, Robert J Marks II (1988). An artificial neural network for spatio-temporal bipolar patterns: Application to phoneme classification. Advances in Neural Information Processing Systems 1: 31–40.
35. D H Hubel, T N Wiesel (1962). Receptive fields, binocular interaction and functional architecture in the cat's visual cortex. J Physiol 160(1): 106–154.
36. B S Hansson, E Hallberg, C Löfstedt, R A Steinbrecht (1994). Correlation between dendrite diameter and action potential amplitude in sex pheromone specific receptor neurons in male *Ostrinia nubilalis* (Lepidoptera: Pyralidae). Tissue Cell 26: 503–512.
37. Alec Radford, Luke Metz, Soumith Chintala (2016). Unsupervised representation Learning with deep convolutional generative adversarial networks. ICLR2016. DOI: 10.48550/arXiv.1511.06434
38. Ryuta Mizutani, Senta Noguchi, Rino Saiga, Yuichi Yamashita, Mitsuhiro Miyashita, Makoto Arai, Masanari Itokawa (2022). Schizophrenia-mimicking layers outperform conventional neural network layers. Front Neurorobot 16: 851471.
39. Yunjey Choi, Youngjung Uh, Jaejun Yoo, Jung-Woo Ha (2020). StarGAN v2: Diverse image synthesis for multiple domains. 2020 IEEE/CVF Conference on Computer Vision and Pattern Recognition (CVPR), Seattle, WA, USA, 2020, pp 8185–8194. DOI: 10.1109/CVPR42600.2020.00821
40. Ziwei Liu, Ping Luo, Xiaogang Wang, Xiaoou Tang (2015). Deep learning face attributes in the wild. 2015 IEEE International Conference on Computer Vision (ICCV), Santiago, Chile, pp 3730–3738. DOI: 10.1109/ICCV.2015.425
41. Martin Heusel, Hubert Ramsauer, Thomas Unterthiner, Bernhard Nessler, Sepp Hochreiter (2017). GANs trained by a two time-scale update rule converge to a local Nash equilibrium. NIPS'17: Proceedings of the 31st International Conference on Neural Information Processing Systems December 2017. pp 6629–6640.
42. Mehdi Mirza, Simon Osindero (2014). Conditional Generative Adversarial Nets. DOI: 10.48550/arXiv.1411.1784
43. Olaf Ronneberger, Philipp Fischer, Thomas Brox (2015). U-Net: Convolutional networks for biomedical image segmentation. In: N Navab, J Hornegger, W Wells, A Frangi (eds) Medical Image Computing and Computer-Assisted Intervention – MICCAI 2015. Lecture Notes in Computer Science, vol. 9351. Springer. pp 234–241.
44. Jiaming Song, Chenlin Meng, Stefano Ermon (2021). Denoising diffusion implicit models. ICLR 2021. DOI: 10.48550/arXiv.2010.02502

45. Pablo F Alcantarilla, Jesús Nuevo, Adrien Bartoli (2011). Fast explicit diffusion for accelerated features in nonlinear scale spaces. Trans Pattern Anal Machine Intell 34(7): 1281–1298.

46. C S Roy, C S Sherrington (1890). On the regulation of the blood-supply of the brain. J Physiol 11(1-2): 85–158.

47. Luc Leybaert (2005). Neurobarrier coupling in the brain: a partner of neurovascular and neurometabolic coupling? J Cereb Blood Flow Metab 25(1): 2–16.

48. Rino Saiga, Masayuki Uesugi, Akihisa Takeuchi, Kentaro Uesugi, Yoshio Suzuki, Susumu Takekoshi, Chie Inomoto, Naoya Nakamura, Youta Torii, Itaru Kushima, Shuji Iritani, Norio Ozaki, Kenichi Oshima, Masanari Itokawa, Makoto Arai, Ryuta Mizutani (2021). Brain capillary structures of schizophrenia cases and controls show a correlation with their neuron structures. Sci Rep 11(1): 11768.

49. B Zwaagstra, D Kernell (1981). Sizes of soma and stem dendrites in intracellularly labelled alpha-motoneurones of the cat. Brain Res 204(2): 295–309.

50. John C Fiala, Kristen M Harris (1999). Dendrite structure. In Dendrites (eds. Greg Stuart, Nelson Spruston, Michael Häusser). Oxford University Press.

51. Blanca Paramo, Sven O Bachmann, Stéphane J Baudouin, Isabel Martinez-Garay, Alun M Davies (2021). Neuregulin-4 is required for maintaining soma size of pyramidal neurons in the motor cortex. eNeuro 8(1): ENEURO.0288-20.2021.

52. Garilyn M Jentarra, Shannon L Olfers, Stephen G Rice, Nishit Srivastava, Gregg E Homanics, Mary Blue, Sakkubai Naidu, Vinodh Narayanan (2010). Abnormalities of cell packing density and dendritic complexity in the MeCP2 A140V mouse model of Rett syndrome/X-linked mental retardation. BMC Neurosci 11: 19.

53. Sampathkumar Rangasamy, Shannon Olfers, Brittany Gerald, Alex Hilbert, Sean Svejda, Vinodh Narayanan (2016). Reduced neuronal size and mTOR pathway activity in the Mecp2 A140V Rett syndrome mouse model. F1000Res 5: 2269.

54. M Botvinick, L E Nystrom, K Fissell, C S Carter, J D Cohen (1999). Conflict monitoring versus selection-for-action in anterior cingulate cortex. Nature 402: 179–181.

55. G Bush, P Luu, M I Posner (2000). Cognitive and emotional influences in anterior cingulate cortex. Trends Cogn Sci 4: 215–222.

56. Paul G Anastasiades, Adam G Carter (2022). Circuit organization of the rodent medial prefrontal cortex. Trends Neurosci 44(7): 550–563.

57. Erhan Genç, Christoph Fraenz, Caroline Schlüter, Patrick Friedrich, Rüdiger Hossiep, Manuel C Voelkle, Josef M Ling, Onur Güntürkün, Rex E Jung (2018). Diffusion markers of dendritic density and arborization in gray matter predict differences in intelligence. Nat Commun 9(1): 1905.

58. Y Suzuki, K Uesugi, N Takimoto, T Fukui, K Aoyama, A Takeuchi, H Takano, N Yagi, T Mochizuki, S Goto, K Takeshita, S Takahashi, H Ohashi, Y Furukawa, T Ohata, T Matsushita, Y Ishizawa, H Yamazaki, M Yabashi, T Tanaka, H Kitamura, T Ishikawa (2004). Construction and commissioning of a 248 m-long beamline with X-ray undulator light source. AIP Conference Proceedings 705: 344–347.

59. A Takeuchi, K Uesugi, H Takano, Y Suzuki (2002). Submicrometer-resolution three-dimensional imaging with hard x-ray imaging microtomography. Rev Sci Instrum 73: 4246–4249.

60. Y Suzuki, A Takeuchi, Y Terada, K Uesugi, R Mizutani (2016). Recent progress of hard x-ray imaging microscopy and microtomography at BL37XU of SPring-8. AIP Conference Proceedings 1696: 020013.

61. A Takeuchi, K Uesugi, Y Suzuki (2009). Zernike phase-contrast x-ray microscope with pseudo-Kohler illumination generated by sectored (polygon) condenser plate. J Phys Conf Ser 186: 012020.

62. V De Andrade, A Deriy, M J Wojcik, D Gürsoy, D Shu, K Fezzaa, F De Carlo (2016). Nanoscale 3D imaging at the Advanced Photon Source. SPIE Newsroom. DOI: 10.1117/2.1201604.006461

63. V De Andrade, V Nikitin, M Wojcik, A Deriy, S Bean, D Shu, T Mooney, K Peterson, P Kc, K Li, S Ali, K Fezzaa, D Gürsoy (2021). Fast X-ray nanotomography with sub-10 nm resolution as a powerful imaging tool for nanotechnology and energy storage applications. Adv Mater 33: e2008653.

64. R Mizutani, R Saiga, S Takekoshi, C Inomoto, N Nakamura, M Itokawa, M Arai, K Oshima, A Takeuchi, K Uesugi, Y Terada, Y Suzuki (2016). A method for estimating spatial resolution of real image in the Fourier domain. J Microsc 261: 57–66.

65. R Mizutani, A Takeuchi, K Uesugi, Y Suzuki (2008). Evaluation of the improved three-dimensional resolution of a synchrotron radiation computed tomograph using a micro-fabricated test pattern. J Synchrotron Radiat 15: 648–654.

66. R Mizutani, A Takeuchi, K Uesugi, S Takekoshi, R Y Osamura, Y Suzuki (2010). Microtomographic analysis of neuronal circuits of human brain. Cereb Cortex 20: 1739–1748.

67. R Mizutani, R Saiga, A Takeuchi, K Uesugi, Y Suzuki (2013). Three-dimensional network of Drosophila brain hemisphere. J Struct Biol 184: 271–279.

68. Sergey Ioffe, Christian Szegedy (2015). Batch normalization: Accelerating deep network training by reducing internal covariate shift. DOI: 10.48550/arXiv.1502.03167

69. Takeru Miyato, Toshiki Kataoka, Masanori Koyama, Yuichi Yoshida (2018). Spectral normalization for generative adversarial networks. ICLR 2018. DOI:

10.48550/arXiv.1802.05957

70. Diederik P Kingma, Jimmy Ba (2015). Adam: A method for stochastic optimization. The 3rd International Conference for Learning Representations, San Diego, 2015. DOI: 10.48550/arXiv.1412.6980

71. Yuxin Wu, Kaiming He (2018). Group normalization. DOI: 10.48550/arXiv.1803.08494

72. Arajit Ramachandran, Barret Zoph, Quoc V Le (2017). Swish: A self-gated activation function. DOI: 10.48550/arXiv.1710.05941

**Supplementary Table S5. (A)** GAN generator using mouse-mimetic convolutional layers. FC, fully connected layer; mConv, mouse-mimetic 2-dimensional convolutional layer using a kernel size of 5 × 5, stride of 2, and the same padding; Conv, standard 2-dimensional convolutional layer using a kernel size of 5 × 5, stride of 2, and the same padding. The Number-of-parameters column shows the number of trainable parameters before the parameter reduction.

| Layer | Output size | Filter size | Number of parameters | Options |
|---|---|---|---|---|
| Latent input | 100 | | | |
| FC | 16384 (4 × 4 × 1024) | | 1,654,784 | |
| Batch norm. | | | 65,536 | Activation: ReLU |
| mConv | 8 × 8 × 512 | 512 | 13,107,712 | |
| Batch norm. | | | 2,048 | Activation: ReLU |
| mConv | 16 × 16 × 256 | 256 | 3,277,056 | |
| Batch norm. | | | 1,024 | Activation: ReLU |
| mConv | 32 × 32 × 128 | 128 | 819,328 | |
| Batch norm. | | | 512 | Activation: ReLU |
| Conv | 64 × 64 × 3 | 3 | 9,603 | Activation: sigmoid |

**Supplementary Table S5. (B)** GAN discriminator configuration. Conv, standard 2-dimensional convolutional layer using a kernel size of 4 × 4, stride of 2, and the same padding; FC, fully connected layer.

| Layer | Output size | Filter size | Number of parameters | Options |
|---|---|---|---|---|
| Input | 64 × 64 RGB | | | |
| Conv | 32 × 32 × 64 | 64 | 3,136 | |
| Spectral norm. | | | 64 | Activation: leaky ReLU 0.2 |
| Conv [1] | 16 × 16 × 128 | 128 | 131,200 | |
| Spectral norm. | | | 128 | Activation: leaky ReLU 0.2 |
| Conv [1] | 8 × 8 × 256 | 256 | 524,544 | |
| Spectral norm. | | | 256 | Activation: leaky ReLU 0.2 |
| Conv [1] | 4 × 4 × 512 | 512 | 2,097,664 | |
| Spectral norm. | | | 512 | Activation: leaky ReLU 0.2 |
| FC | 1 | | 8193 | Dropout: 20% Activation: sigmoid |

[1] When examining mouse-mimetic convolutional layers in the discriminator, these Conv layers were replaced with their mouse-mimetic versions.

**Supplementary Table S6.** Number of weights and %usage in the GAN generator. mConv, mouse-mimetic 2-dimensional convolutional layer; Conv, standard 2-dimensional convolutional layer.

| Distance threshold [1] | 0.2 | 0.3 | 0.4 | 0.6 | 0.8 | 1.4 |
|---|---|---|---|---|---|---|
| Number of weights / layer [2] | | | | | | |
|   Fully connected | 1,638,400 | 1,638,400 | 1,638,400 | 1,638,400 | 1,638,400 | 1,638,400 |
|   mConv | 1,384,200 | 2,846,150 | 4,481,150 | 8,106,750 | 11,151,550 | 13,107,200 |
|   mConv | 344,400 | 709,600 | 1,119,125 | 2,023,450 | 2,784,475 | 3,276,800 |
|   mConv | 98,350 | 169,400 | 284,200 | 512,550 | 693,950 | 819,200 |
|   Conv | 9,600 | 9,600 | 9,600 | 9,600 | 9,600 | 9,600 |
| Total number of weights | | | | | | |
|   Overall | 3,474,950 | 5,373,150 | 7,532,475 | 12,290,750 | 16,277,975 | 18,851,200 |
|   Mouse layers | 1,826,950 | 3,725,150 | 5,884,475 | 10,642,750 | 14,629,975 | 17,203,200 |
| %Usage of weights | | | | | | |
|   Overall | 18.4 | 28.5 | 40.0 | 65.2 | 86.3 | 100.0 |
|   Mouse layers | 10.6 | 21.7 | 34.2 | 61.9 | 85.0 | 100.0 |

[1] Distance in fractional coordinate.

[2] Layers in the GAN generator shown in Supplementary Table S5A.

**Supplementary Table S8.** Number of weights and %usage in the U-Net of DDIM.

| Distance threshold [1] | 0.4 | 1.4 |
|---|---|---|
| Total number of used weights | | |
|   Overall | 7,965,134 | 18,095,552 |
|   Mouse layers | 5,352,462 | 15,482,880 |
| %Usage of weights | | |
|   Overall | 44.0 | 100.0 |
|   Mouse layers | 34.6 | 100.0 |

[1] Distance in fractional coordinate.

**Supplementary Table S10.** Folders of the Cheese Pics dataset used in this study.

| abbaye de belloc cheese | abondance cheese | aged gouda cheese |
|---|---|---|
| allgauer emmentaler cheese | alpicreme cheese | ambert cheese |
| ameribella cheese | appenzeller cheese | applebys double gloucester cheese |
| asiago cheese | asiago pressato cheese | baby swiss cheese |
| baita friuli cheese | barely buzzed cheese | baron bigod cheese |
| bayley hazen blue cheese | beaufort cheese | beemster aged cheese |
| beemster classic cheese | beemster extra aged cheese | beenleigh blue cheese |
| bethmale des pyrenees cheese | bleu dauvergne cheese | bleu de laqueuille cheese |
| bleu des causses cheese | blue vein australian cheese | blue vein cheese cheese |
| brebis du lavort cheese | brebis du puyfaucon cheese | brie de meaux cheese |
| brie de melun cheese | cabot clothbound cheese | caerphilly cheese |
| canestrato cheese | cantal cheese | challerhocker cheese |
| chaumes cheese | cheddar cheese | chorlton blue cheshire cheese |
| colston bassett stilton cheese | consider bardwell farm manchester cheese | coolea cheese |
| coquetdale cheese | cornish crumbly cheese | cornish kern cheese |
| cornish yarg cheese | coulommiers cheese | cows milk gouda cheese |
| cravero parmigiano reggiano cheese | crayeux de roncq cheese | danish fontina cheese |
| devon blue cheese | dorset cheese | dorset blue vinny cheese |
| double gloucester cheese | dry jack cheese | emental grand cru cheese |
| emmental cheese | etorki cheese | ewes blue cheese |
| fondant de brebis cheese | fontina cheese | fontina val daosta cheese |

**Supplementary Table S10** (cont'd). Folders of the Cheese Pics dataset used in this study.

| four herb gouda cheese | fourme de montbrison cheese | gippsland blue cheese |
|---|---|---|
| gloucester cheese | gloucester goat cheese | gorgonzola cheese |
| gorgonzola dolce dop cheese | gorwydd caerphilly cheese | gouda cheese |
| grana cheese | grana padano cheese | grayson cheese |
| hafod cheese | harbourne blue cheese | huntsman cheese |
| jarlsberg cheese | kadchgall cheese | kasseri cheese |
| keens cheddar cheese | kirkhams lancashire cheese | kolan extra mature cheese |
| la couronne fort aged comte cheese | la peral cheese | lairobell cheese |
| lamuse signature gouda cheese | lanark blue cheese | landaff cheese |
| langres cheese | lappi cheese | le brin cheese |
| le gruyere aop cheese | leyden cheese | liliputas cheese |
| lincolnshire poacher cheese | lingot des causses cheese | livarot cheese |
| longhorn cheese | maasdam cheese | maffra red leicester cheese |
| mahoe aged gouda cheese | mahon cheese | manchego cheese |
| milawa affine cheese | milawa goats tomme cheese | mimolette boule de lille cheese |
| mobay cheese | montagnolo cheese | montasio cheese |
| montasio mezzano cheese | monte enebro cheese | monteo cheese |
| montgomerys cheddar cheese | morbier cheese | mt tam cheese |
| munster cheese | nettle meadow kunik cheese | oak smoked cheddar cheese |

**Supplementary Table S10** (cont'd). Folders of the Cheese Pics dataset used in this study.

| ogleshield cheese | old goat cheese | old winchester cheese |
|---|---|---|
| oregon blue cheese cheese | organic gouda cheese | ossau fermier cheese |
| pecorino al pepe cheese | pecorino al tartufo cheese | pecorino cheese |
| pecorino pepato cheese | pepato cheese | pleasant ridge reserve cheese |
| pont leveque cheese | reblochon cheese | roquefort cheese |
| saint paulin cheese | sheep gouda cheese | shropshire blue cheese |
| somerset brie cheese | spenwood cheese | stella kasseri cheese |
| stichelton cheese | stilton cheese | stinking bishop cheese |
| swiss cheese | taleggio cheese | tarentaise cheese |
| teahive cheese | ticklemore cheese | toma cheese |
| toma di campo cheese | toma piemontese cheese | tomme cheese |
| tomme de chevre cheese | tomme de romans cheese | tomme de savoie cheese |
| tommes cheese | tomme des chouans cheese | truffle tremor chees |
| twig farm washed rind wheel cheese | vacherin fribourgeois cheese | vento destate cheese |
| washed rind cheese australian cheese | west country farmhouse mature cheddar cheese | wookey hole cave aged cheddar cheese |

Other supplemental tables and figures will be provided upon journal publication.